\PassOptionsToPackage{mathlines}{lineno}
\documentclass[twocolumn,iop,apj]{aastex631} 
\usepackage{times}
\usepackage[sort&compress]{natbib}

\usepackage{amssymb,amsmath,array} 
\usepackage{graphicx,subfigure}
\usepackage{color}
\usepackage{ulem} 
\usepackage{float}
\usepackage{listings}
\usepackage{xparse}

\newcolumntype{F}[1]{>{\minipage{\dimexpr#1}\centering\arraybackslash}c<{\vspace\tabcolsep\endminipage}}

\newcommand{\be}{\begin{equation}}
\newcommand{\ee}{\end{equation}}

\newcommand{\hi}{H~{\sc i}}

\newcommand{\hii}{H~{\sc ii}}
\newcommand{\e}[2]{$^{#1}_{#2}$}

\newcommand{\galfit}{{\sc GalFit}}

\newcommand{\GP}{{\sc GalProp}}

\newcommand{\galprop}{{\sc GalProp}}

\shorttitle{Cosmic Rays From the Nearby ISM}
\shortauthors{Porter~et~al.}

\begin{document}



\title{Voyager~1 Data Reveals Signatures of the Local Gas and Cosmic-Ray Source Distributions}
\correspondingauthor{T.~A.~Porter}
\email{tporter@stanford.edu}
\author[0000-0002-2621-4440]{T. A. Porter}
\affil{Hansen Experimental Physics Laboratory, Stanford University, Stanford, CA 94305, USA}
\affil{ Kavli Institute for Particle Astrophysics and Cosmology, Stanford University, Stanford, CA 94305, USA}

\author[0000-0001-6141-458X]{I. V. Moskalenko}
\affil{Hansen Experimental Physics Laboratory, Stanford University, Stanford, CA 94305, USA}
\affil{ Kavli Institute for Particle Astrophysics and Cosmology, Stanford University, Stanford, CA 94305, USA}

\author[0000-0002-3840-7696]{A. C. Cummings}
\affiliation{California Institute of Technology, Pasadena, CA 91125, USA}


\author[0000-0003-1458-7036]{G. J{\'o}hannesson}
\affil{Science Institute, University of Iceland, Dunhaga 3, IS-107 Reykjavik, Iceland}

\begin{abstract}
  We investigate the effects of the nearby interstellar medium (ISM) on the locally measured cosmic‑ray (CR) spectra.
  Using the \GP\ code we explore how variations in the local gas and source distributions affect spectral features at low energies. 
  Comparing with recent Voyager~1 measurements taken in the local ISM, we show that for a realistic interstellar gas distribution the data favour models in which there are no significant CR sources within $\sim$150-200~pc of the Solar system, implying that the nearest dominant contributors to the low-energy CR flux are located at distances beyond this range.
  We find that the modelling supports the conclusion of \citet{Cummings_2025} that there is a significant fraction of primary Boron in its observed spectrum at low energies.
  Our study shows that detailed modelling of the immediate Galactic environment is required to robustly infer Galactic CR propagation parameters from local measurements, and that accounting for nearby ISM structure can alleviate tensions between direct CR data and global propagation models.
\end{abstract}



\section{Introduction} \label{sec:intro}
\setcounter{footnote}{0}

Cosmic rays (CRs) are a uniquely sensitive probe of the local interstellar medium (ISM) conditions.
The combined influence of nearby CR sources, small-scale variations in the interstellar gas, and the dynamical history of the Solar neighbourhood are encoded in their spectra at both low ($\lesssim$1~GeV) \citep[e.g.,][and references therein]{2016Sci...352..677B,2021ApJ...913....5B} and very high ($\gtrsim$1~TeV) energies \citep[e.g.,][and references therein]{2024AdSpR..74.4264M}.
For decades, attempts to access the low-energy signatures have been hampered by the modulation of CR spectra by the Solar wind.
But measurements by Voyager 1 (V1) beyond the heliopause now offer the first direct view of the local interstellar spectrum down to a few MeV nucleon$^{-1}$ energies \citep{Cummings_2025}.
These data open a window onto CR transport on scales of tens to hundreds of parsecs, which are at the scales where the structures of the very local ISM (VLISM) are known to vary considerably compared to averaging over the broader Galactic disc.

A defining feature of the nearby Galactic environment is the Local Bubble (LB): a cavity of hot, tenuous plasma $\sim$50–200~pc around the Solar system, bounded by a complex shell of neutral and molecular material.
Recent 3D reconstructions of dust and gas reveal that this cavity is highly irregular \citep[e.g.,][]{2024ApJ...973..136O,2025A&A...695A.222S} and was likely formed by a sequence of supernova (SN) explosions over the past 10--20~Myr \citep{2016Natur.532...73B,Zucker_2022}.
The LB’s very low gas density modifies the dominant loss processes for low-energy CRs, while its apparent lack of recent SN activity implies a deficit of local CR injection.
These are coupled consequences of the same evolutionary history, which have so far not been treated together in a self-consistent modelling framework.


In this paper, we use the \GP\ code to quantify how realistic gas distributions and possible local deficits in CR sources affect the spectra observed by V1.
Our 3D modelling uses high-resolution, data-driven gas reconstructions \citep{2025A&A...693A.139S} that capture the LB's underdense interior and its surrounding gas structures.
We also test the impact of a spherical deficit (``hole'') in the CR source distribution centred on the Solar system, representing the lack of recent nearby SN explosions in the local ISM.
Each 3D configuration is tuned to match a two-dimensional (2D) baseline model \citep{Cummings_2025} at energies $\gtrsim$few 100 MeV nucleon$^{-1}$ obtained from a fit to the V1 data, ensuring that deviations at lower energies reflect local structure rather than global transport differences.

We show that neither the gas structure nor the source distribution alone is sufficient to reproduce the V1 low-energy spectra.
Instead, the data require a simultaneous local deficit in both gas and sources, consistent with the likely physical properties and formation history of the LB. In particular, for realistic inhomogeneous gas distribution the nearest significant CR source contributing to the low-energy flux must lie at $\sim$150--200 pc from the Solar system.
This is comparable to the LB’s extent and consistent with CRs in the VLISM coming from injection near its periphery.

\section{\GP\ Framework}
The \GP{} package is a state-of-the-art comprehensive framework for analysis and interpretation of the measurements of CRs and their emissions.
Theoretical understanding of CR propagation in the ISM is the framework that the \GP{} code is built around.
The key idea is that all CR-related data, including direct measurements, $\gamma$-ray{s}, synchrotron radiation, etc., are subject to the same physics and must therefore be modelled self-consistently \citep{1998A&A...338L..75M}.
\GP{} can combine the results of individual measurements in physics and astronomy spanning in energy coverage, types of instrumentation, and the nature of detected species. Its range of physical validity extends from sub-keV$-$PeV energies for particles and from $10^{-6}$ eV ($\mu$eV)$-$PeV for photons.
The goal for the \GP{}-based models is to be as realistic as possible and to make use of all available astrophysical information, nuclear and particle data, with a minimum of simplifying assumptions \citep{2007ARNPS..57..285S}.

\GP{} has 27+ years of development behind it \citep{1998ApJ...493..694M, 1998ApJ...509..212S}.
It is consistently updated to keep up with the ever increasing amount and precision of experimental data.
The latest release of \galprop{}, version 57.1, became public in 2022 \citep[][and references therein]{2022ApJS..262...30P}.
The dedicated website\footnote{https://galprop.stanford.edu \label{site}} provides source releases, supporting data products and run configurations for reproducing published results, and a facility to run \GP\ releases via web browsers \citep{2011CoPhC.182.1156V}.

The \GP{} code solves a system of about 90+ time-dependent transport equations (partial differential equations in 3D or 4D: spatial variables plus energy) with a given source distribution and boundary conditions to give the intensity distributions for all CR species through the ISM: $^1$H$-$\,$^{64}$Ni, $\bar{p}$, $e^\pm$. 
The spatial boundary conditions assume free particle escape. 
The propagation equations include terms for convection, distributed reacceleration, energy losses, nuclear fragmentation, radioactive decay, and production of secondary particles and isotopes \citep[for details of these processes and formalism, see][]{1998ApJ...509..212S}. 

For a given halo size, the diffusion coefficient $D_{xx}(\rho)$, as a function of rigidity $\rho$, and other propagation parameters can be determined from secondary-to-primary nuclei ratios, typically B/C, [Sc+Ti+V]/Fe, and/or $\bar{p}/p$. 
If reacceleration is included, the momentum-space diffusion coefficient $D_{pp}$ is related to the spatial coefficient $D_{xx} = \beta D_0 \rho^\delta$ \citep{1994ApJ...431..705S}, where $\beta=v/c$ is the particle velocity, $\rho$ is the magnetic rigidity, and $\delta = 1/3$ for a Kolmogorov spectrum of interstellar turbulence \citep{1941DoSSR..30..301K}, or $\delta = 1/2$ for an Iroshnikov$-$Kraichnan cascade \citep{1964SvA.....7..566I, 1965PhFl....8.1385K}, but can also be arbitrary.
The spatial diffusion coefficient can also depend on position. 
This option was developed and utilised by \citet{2015ApJ...799...86A}, where the positional dependence of the spatial diffusion coefficient was linked to the distribution of the Galactic magnetic field (GMF) strength.

CR source distributions can be specified using a composition scheme that allows the spatial density distribution, spectral characteristics, and respective contributions to be customised.
Possible components for the spatial density model include an axisymmetric disc, spiral arms, various central bulges, and other structures.
  Each basic component can be further split up and fine-tuned with different radial profiles, so that different classes of sources have their own population spatial distribution, injection spectra, and isotopic abundances, allowing for a very flexible description of a galaxy.

The injection spectra of CR species for a source density distribution are parameterised by a multiple broken power law in rigidity:
\begin{equation}
q(\rho) \propto (\rho/\rho_0)^{-\gamma_0}\prod_{i=0}^N\left[1 + (\rho/\rho_i)^\frac{\gamma_i - \gamma_{i+1}}{s_i}\right]^{s_i},
\label{eq:injection}
\end{equation}
where $\gamma_{i =0,\dots,N+1}$ are the spectral indices, $\rho_{i = 0,\dots,N}$ are the break rigidities, and $s_i$ are the smoothing parameters ($s_i$ is negative/positive for $|\gamma_i |\lessgtr |\gamma_{i+1} |$). Each primary isotope can have unique spectral parameters. 

The \GP{} code computes a complete network of primary, secondary, and tertiary isotope production starting from input CR source abundances.
The nuclear reaction network is built using the \citet{2018NDS...151D...3.}, where a detailed description of its method of construction is given by \citet{2020ApJS..250...27B,BoDe21}. 
Included are multistage chains of $p$-, $n$-, $d$-, $t$-, $^3$He-, $\alpha$-, and $\beta^\pm$-decays, and K-electron capture, as well as, in several cases, more complicated reactions.
This accounts for up to 81 daughter nuclei in the final state for each fragment produced in the spallation of the target nucleus, plus an unlimited number of $p$-, $n$-, and $\beta^\pm$-decays. 

Because the decay branching ratios and half-lives of fully stripped and hydrogen-like ions may differ (a well-known example is $^7$Be), \galprop{} includes the processes of K-electron capture, electron pickup from the neutral ISM gas, and formation of hydrogen-like ions as well as the inverse process of electron stripping \citep{1973RvMP...45..273P, 1978PhDT........12W, 1979PhDT........67C}. Meanwhile, the fully stripped and hydrogen-like ions are treated as separate species.

The isotopic production cross section routines are built using LANL nuclear codes \citep{2001ICRC....5.1836M, 2003ICRC....4.1969M, 2004AdSpR..34.1288M, 2005AIPC..769.1612M, 1996PhRvC..54.1341B, 1998PhRvC..57..233B}, databases (LANL, EXFOR\footnote{https://www-nds.iaea.org/exfor/}), an extensive literature search \citep{BoDe21}, and parameterizations \citep{2003ApJS..144..153W, 1998ApJ...501..911S, 1998ApJ...501..920T, 1983ApJS...51..271L}. 
Cross sections for production of $^2$H, $^3$H, $^3$He, and total inelastic cross sections parameterizations \citep{1996PhRvC..54.1329W, BarPol1994} and corrected Tripathi formalism are described in \citet[][Appendices F, G]{2022ApJS..262...30P}. 
The latest updates of the production cross sections are summarised in \citet[][Appendix B.4]{Cummings_2025}.

For the CR interactions with the interstellar gas, \galprop{} runs can use different density models. 
The ISM gas consists mostly of H and He with a ratio of 10:1 by number \citep{2001RvMP...73.1031F}.
Hydrogen can be found in the different states, atomic (\hi), molecular (H$_2$), or ionised (\hii), while He is mostly neutral.
\hi{} is $\sim$60\% of the mass, while H$_2$ and \hii{} contain 25\% and 15\%, respectively \citep{2001RvMP...73.1031F}.
The \hii{} gas has a low number density and scale height $\sim$few~100~pc.
The H$_2$ gas is clumpy and forms high-density molecular clouds.

For 2D calculations, analytical models for the gas density distribution are available \citep{2002ApJ...565..280M} (GP2D).
The radial distribution for \hi{} is taken from \citet{1976ApJ...208..346G} while the vertical distribution is from \citet{1990ARA&A..28..215D} for galactocentric radial distances $0\le R\le 8$~kpc and \citet{1986A&A...155..380C} for $R\ge10$~kpc with linear interpolation in between.
The CO gas distribution is taken from \citet{1988ApJ...324..248B} for 1.5~kpc$<$$R$$<$10~kpc, and from \citet{1990A&A...230...21W} for $R$$\ge$10~kpc, and is augmented with the \cite{2007A&A...467..611F} model for $R$$\le$1.5 kpc.
Both of the 2D atomic and molecular gas distributions are rescaled to the common IAU recommended Sun-Galactic centre (GC) distance of $R_\odot = 8.5$~kpc \citep{1986MNRAS.221.1023K}.
The \hii{} gas distribution is given by the NE2001 model \citep{2002astro.ph..7156C, 2003astro.ph..1598C, 2004ASPC..317..211C} with the updates from \citet{2008PASA...25..184G}.

For 3D simulations, the \hi\ and $^{12}$CO distributions from \cite{2018ApJ...856...45J} (J18) are available.
These were developed using a maximum-likelihood forward-folding optimisation applied to the LAB-\hi\ \citep{2005A&A...440..775K} and CfA composite CO data \citep{2001ApJ...547..792D, 2004ASPC..317...66D}.
New with this paper we also include the gas distribution from \citet{2025A&A...693A.139S} (S25) who made a reconstruction of a very similar collection of data to the J18 work using 3D Gaussian processes to model correlations in the interstellar gas over different sight lines to enforce coherent structure for its distribution.
Compared to the 2D models, the added degrees of freedom allow the optimised distributions to better reproduce the features observed in the line-emission surveys.

Multiple 2D and 3D models for the interstellar radiation field and GMF are also available with \GP.
However, these affect the propagation/energy losses for the CR electrons/positrons, which are not investigated in the current work.
We therefore refer the interested reader to \citet{2022ApJS..262...30P} and associated references for more details.

\section{Calculations}

\subsection{Model Configurations}
\begin{figure*}[htb!]
  \includegraphics[scale=0.8]{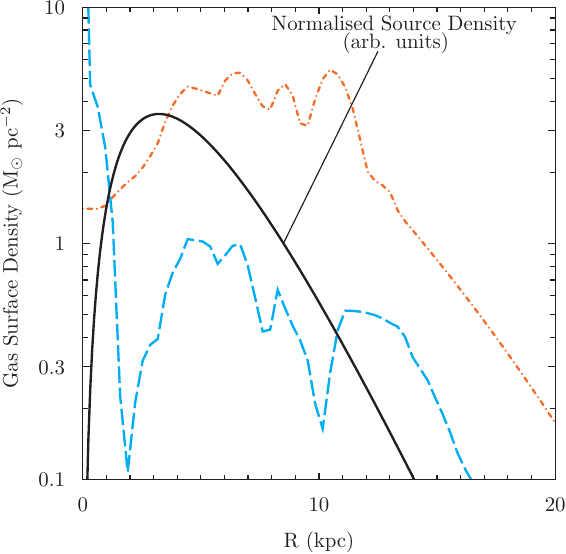}
  \includegraphics[scale=0.8]{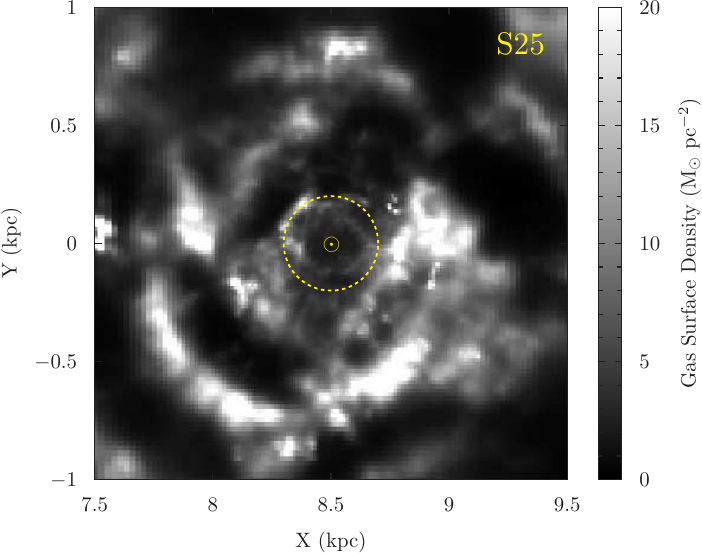}
  \caption{Model configuration components. Left panel: profiles along the $Y = 0$~kpc direction for the CR source (solid line, arbitrary units) and GP2D gas surface densities (\hi, dash-dotted line and H$_2$, long dashed line). 
    Right panel: S25 model gas surface density for the nearby Galactic neighbourhood. The Solar system is marked by the Sun symbol in the centre, and the dashed line is a circle of radius 200~pc about it.\label{fig0}}
\end{figure*}

Without loss of generality, we adopt the 4022a plain diffusion (PD) model determined from fitting to the V1 data for the period when it is in the VLISM \citep{Cummings_2025} as a baseline model.
This configuration also reproduces the local interstellar spectrum from 100~MeV nucleon$^{-1}$ up to $\sim$several GeV nucleon$^{-1}$ obtained using the GALPROP-HelMod framework fitting to the AMS-02 data \citep{2020ApJS..250...27B}.
Throughout, we will use the terms `2D' or `baseline' to refer interchangeably to it.

The baseline model is 2D Galactocentric axisymmetric with a spatial model for the CR sources following the distribution of pulsars from \citet{2006MNRAS.372..777L}.
The injection spectrum for the sources is a power-law in rigidity with a single injection index for all species below a break at 3 GV and a different common index for all species above that.
The interstellar gas distribution is the standard 2D model GP2D described above.
The propagation model uses a single power-law index $\delta$ for the diffusion coefficient $\propto D_0\rho^\delta$, where $D_0$ is the normalisation at 4~GV and $\rho$ is rigidity.
The size of the Galaxy has maximum radius of 20~kpc with a $Z$-boundary (halo size) 4~kpc.

For the 3D models, the PD phenomenology and rigidity dependence for the diffusion coefficient is the same as the baseline.
The size of the Galaxy has a maximum $X/Y$ boundary at $\pm$20~kpc and $Z$-boundary at 4~kpc.
For the gas spatial distribution we use the GP2D and S25 models described above.
The molecular gas density for the S25 model is obtained using the baseline/GP2D $X_{CO}$ distribution. 
The source density model is the same as the baseline.
For the 3D solutions we also consider a spherical hole in the source density about the Solar system.
With these models we test the effect of smooth and inhomogeneous source and gas distributions for the nearby Galactic neighbourhood.

For these configurations, we will use the term `3D' to refer to the equivalent \GP\ solution in 3D for the 2D case.
We will use the term `S25' to refer to the 3D solution using the S25 gas model.
When a hole in the source density is used for either, we will append its radius in parsecs to the identifier. 
For example, the S25 model with a 200~pc radius spherical hole in the source density is termed `S25/200' with this scheme.
We will use a set of spherical radii for the source density hole size: 0~pc (no hole), 100~pc, 150~pc, and 200~pc, respectively.

Figure~\ref{fig0} shows the components for the source and gas distributions. Left panel shows profiles for the CR source distribution employed for all models and the GP2D gas surface density taken along the $Y=0$~kpc direction.
Right panel shows the gas surface density $\lesssim$1~kpc for the S25 distribution centred on the Solar system.
It has wide variations with clumpy high density and low density regions throughout the local Galactic neighbourhood.
For the S25 distribution, we overlay also a circle of radius 200~pc as a comparative gauge for the extent of the source density holes with respect to the nearby ISM structure.

All model configuration files necessary to reproduce our results will be available as supplementary material from the \GP\ website as is our usual practice.


\subsection{CR Solution Geometry}

\begin{figure*}[htb!]
  \centering
  \includegraphics[scale=1.3]{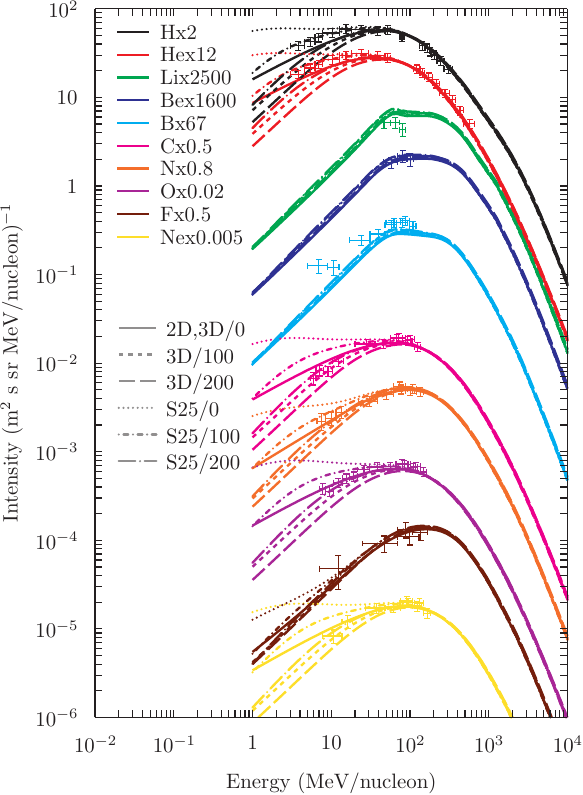}
  \caption{Model combination results together with V1 data for 2013/1--2021/365 in the VLISM from \citet{Cummings_2025} for elemental spectra Z=1--10.
    Line styles: 2D, 3D/0, solid; 3D/100, short dashed; 3D/200, long dashed; S25/0, dotted; S25/100, short dash dotted; S25/200, long dash dotted.\label{fig1a}}
\end{figure*}

For the 2D baseline model, we calculate the CR solution using a spatial grid with linear radial and $Z$ spacing  of 0.25~kpc and 0.1~kpc, respectively.
The kinetic energy grid covers 1--10$^5$~MeV nucleon$^{-1}$ with 15~logarithmic bins/decade spacing.

For 3D models, we calculate the CR solution using a spatial grid with the non-uniform tan grid functions for all coordinates $X,Y,Z$, as described by Appendix~E of \citet{2022ApJS..262...30P}.
The parameters for the individual coordinate functions are chosen so that the pixel size close to the Solar system is $\sim$12~pc.
For the $X/Y$ coordinates the parameters are chosen so that the resolution slowly decreases with distance from the Solar system until it reaches the size of the $Z$ (halo) boundary, and then coarsens more beyond that to the respective planar boundaries.
Because the gas distribution falls rapidly with increasing $Z$, the grid function parameters are chosen so that the resolution slowly decreases only until a distance 1~kpc above the plane and then coarsens more beyond that to the halo boundary.

Our choice of minimum voxel size is guided by initial testing that showed the CR solutions are insensitive to ISM structure below its dimensions. 
The same kinetic energy grid as the 2D baseline is used for all 3D model solutions.
Standard \GP\ modelling 3D geometry adopts a right-handed system with the GC located at (0,0,0) and Solar system on the positive $X$ axis at coordinates ($X,Y,Z$)$=$(8.5,0,0) where distances are in kiloparsecs.

Note that the S25 gas model is derived assuming a Sun-GC distance 8.2~kpc, but is provided in heliocentric coordinates.
For our calculations we place its coordinate origin at the location of the Solar system in our standard setup.
For our grid scheme the voxels $\gtrsim$5~kpc from the Solar system are sufficiently large that the $\sim$0.3~kpc difference between the Sun-GC distances is effectively absorbed.
Because this is beyond the horizon set by the assumed halo size of 4~kpc, there is no effect on our calculations from these Sun-GC distance differences.

\subsection{Tuning Procedure}
\label{sec:tune}
The 2D baseline model was obtained by \citet{Cummings_2025} using the \galfit\ package.
This is a tool that iteratively optimises a \GP\ model configuration parameter set against a user-specfied collection of data.
For the 2D case this is computationally feasible because the likelihood evaluations are very quick, allowing for the parameter space to be adequately sampled.
Such a procedure is not feasible for the high-resolution 3D models in this paper because the computational costs are much higher.


Instead, we take the 2D baseline as a proxy for the data and adjust the 3D configurations to reproduce its spectra for energies $\gtrsim$few GeV/nucleon.
We require that the agreement is within the $\sim$few percent experimental uncertainties on the data that the baseline model was fitted to. 
The motivation for this normalisation method is that global transport effects are generally expected to dominate the shaping of the CR spectra at these energies.

Because we are using a fixed halo size, equivalence for each 3D configuration with the 2D baseline is obtained by tuning parameters when there is no hole in the source spatial distribution.
This amounts to rescaling the diffusion coefficient normalisation so that it reproduces the 2D baseline spectra and B/C at these energies.
For corresponding configurations with the source distribution hole we use the same diffusion coefficient as the no hole case.

\subsection{Results}


\begin{figure*}[htb!]
  \centering
  \includegraphics[scale=1.3]{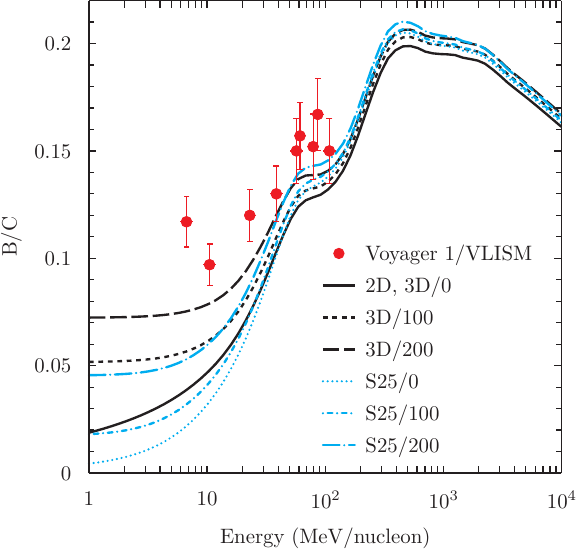}
  \caption{Model combination results together with V1 data for 2013/1--2021/365 in the VLISM from \citet{Cummings_2025} for B/C.
    Line styles as for Fig.~\ref{fig1a}.\label{fig1b}}
\end{figure*}

\begin{figure*}[htb!]
  \includegraphics[scale=0.9]{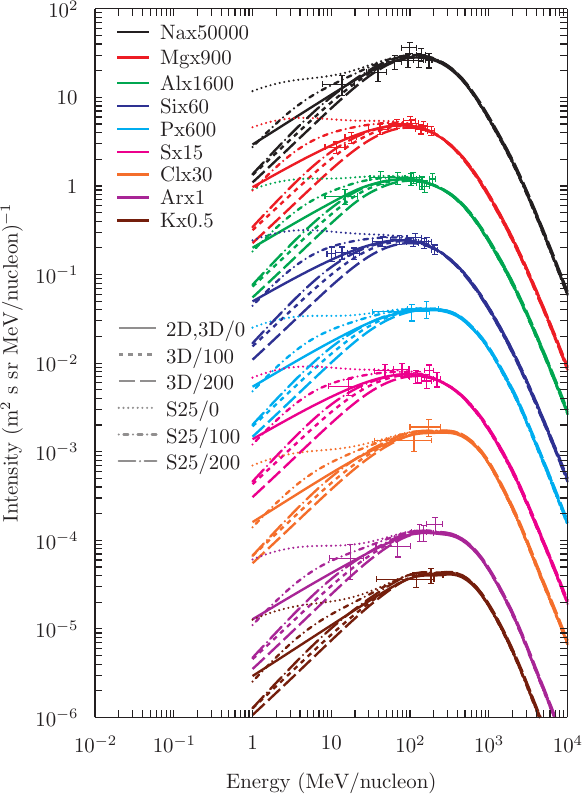}
  \includegraphics[scale=0.9]{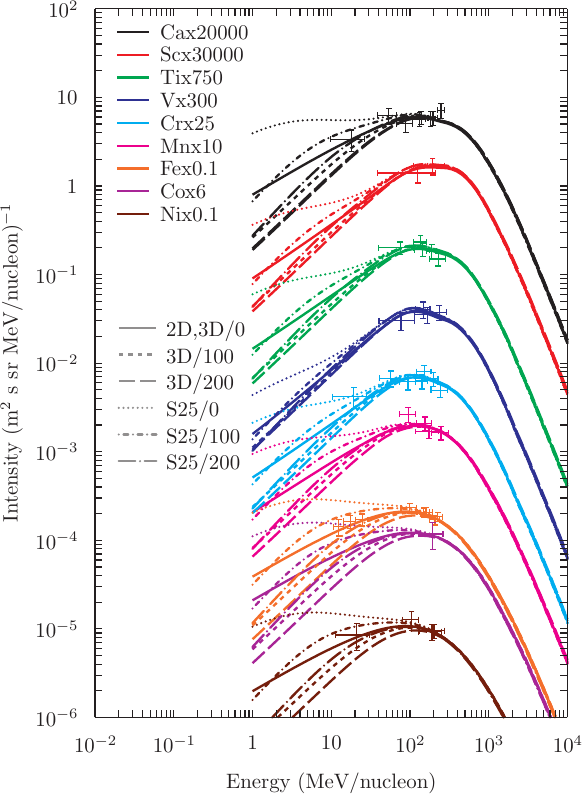}
  \caption{Model combination results together with V1 data for 2013/1--2021/365 in the VLISM from \citet{Cummings_2025} for elemental spectra Z=11--19 (left) and Z=20--28 (right).
    Line styles as for Fig.~\ref{fig1a}.\label{fig2}}
\end{figure*}

\begin{figure*}[htb!]
  \includegraphics[scale=0.9]{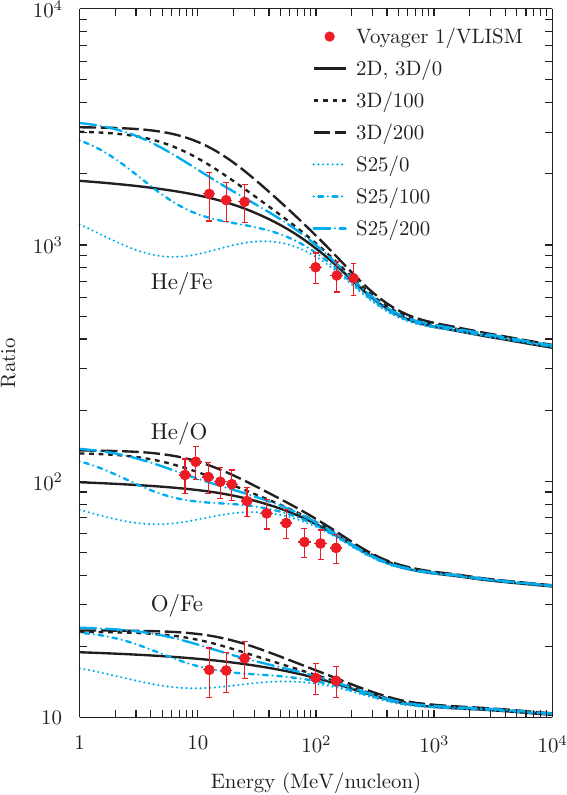}
  \includegraphics[scale=0.9]{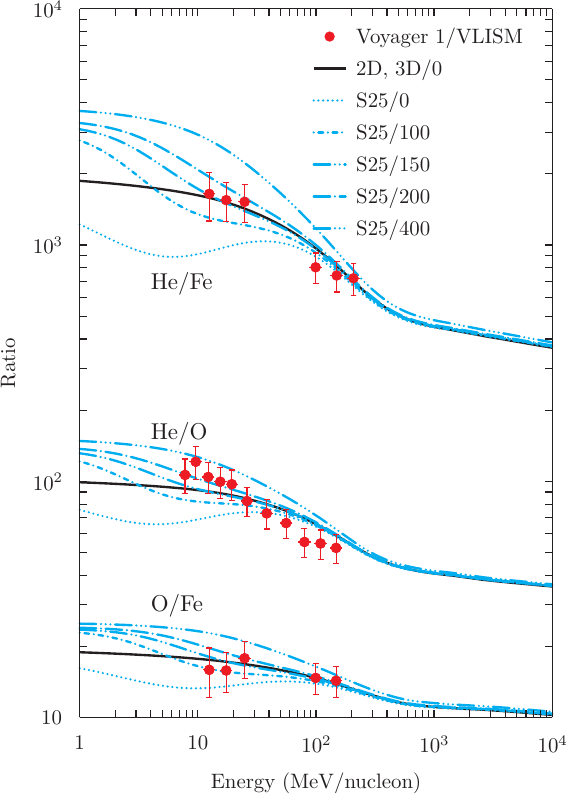}
  \caption{Elemental ratios for He, O, and Fe compared with V1 data. For each ratio, the corresponding V1 ratio has been derived from the spectra using power-law interpolation at the mid-bin energy for the elemental denominator species. Uncertainties are obtained by combining in quadrature the 1-sigma uncertainties for each elemental species interpolated using the same method.
    Left panel: ratios for combinations with 3D/GP2D and S25 gas models with source density hole sizes 0, 100, and 200~pc.
    Right panel: ratios for S25 gas model with source density hole sizes 0, 100, 150, 200, and 400~pc.
    Line styles are same as for Fig.~\ref{fig1a}, with the addition of double-dot dashed and triple-dotted dashed lines for 150~pc and 400~pc source density holes, respectively. \label{fig3}}
\end{figure*}

\begin{figure*}[htb!]
  \centering
  \includegraphics[scale=0.9]{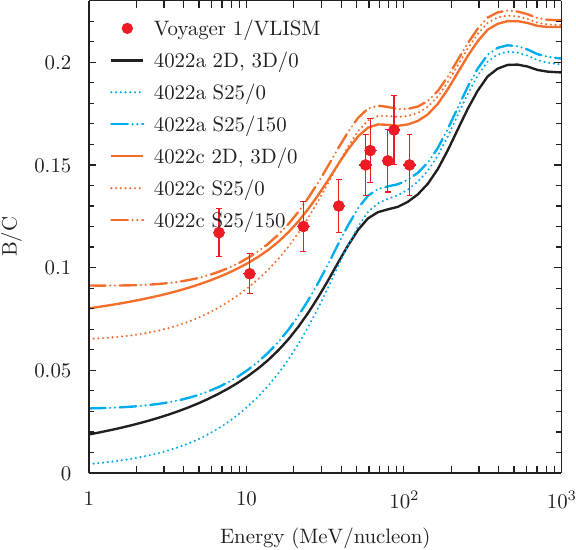}
  \includegraphics[scale=0.9]{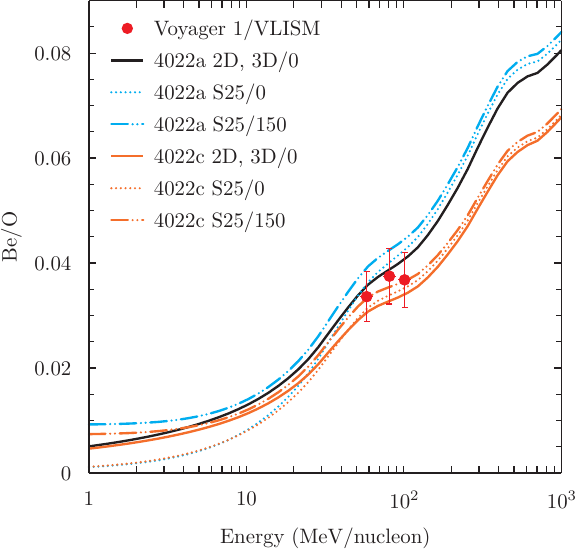}
  \caption{Model results for baseline and S25/150 with Boron pure secondary (4022a) and with primary Boron following 4022c abundances obtained by \citet{Cummings_2025} together with V1 data for 2013/1--2021/365 in the VLISM.
    Panels: B/C, left; Be/O, right.
    The V1 Be/O ratio is derived using power-law interpolation at the geometric mean energy for the elemental Be spectrum.
    Line styles as for Fig.~\ref{fig3}, with orange lines showing 4022c model curves. \label{fig4}}
\end{figure*}

Figure~\ref{fig1a} shows the results for the elemental spectra for Z=1--10 with the V1 data from \citet{Cummings_2025}.
The different line styles correspond to the GP2D and S25 gas models with different source density hole radial sizes: no hole, 100~pc, and 200~pc.
It can be seen that the modelled CR spectra for all configurations closely agree for energies $\gtrsim$100--200~MeV nucleon$^{-1}$ showing consistent normalisation.
Differences occur for energies below this according to the gas model used and source density hole size.

Between the gas models, the CR intensities for the homogeneous (no-hole) source densities are different.
While the GP2D/3D combination is as for the 2D baseline fit, for the S25/0 and S25/100 configurations the low-energy spectra are considerably higher.
The CR injection occurs right up to the immediate vicinity of the Solar system when there is no hole.
The inhomogeneous gas distribution is particularly low ($\lesssim$100~pc) nearby, and this results in slower energy losses.
Consequently, the spectra are not as attenuated as for the homogeneous gas distribution with the GP2D model.

Increasing the source density hole size results in an attenuation of the spectra of predominantly primary species below 100~MeV nucleon$^{-1}$ energies for both gas models.
CRs reaching the Solar system are propagating farther from their initial injection beyond the exclusion region.
Because there is no replenishment over the source density hole, the energy losses during propagation are not counteracted resulting in lower spectra compared to the homogeneous (source) case.

The combination of these effects produces the low-energy behaviour shown in the figure.
For the homogeneous GP2D gas model, a source density hole size $\sim$100~pc in radius gives spectra that remain consistent with the data.
Meanwhile, for the S25 model a source density hole size $\gtrsim$100~pc is necessary, with a likely range $\sim$150--200~pc to obtain reasonable consistency with the data.

For all model configurations the light secondaries Li--Be--B are unaffected by the localised features for the CR or gas distributions.
At high energies this is expected.
But at the lower energies there is no significant effect either.
This is showing that the production of these species, which are pure secondaries for the 4022a scenario abundances that the 2D/baseline employs, is over a larger volume than the fairly nearby localised features (concentrations or deficits) in the gas or CR density or that spectral fluctuations can affect the primary spectra.

Figure~\ref{fig1b} shows the B/C ratio for the same model configurations as in Fig.~\ref{fig1a}.
The ratios over all configurations for energies $\gtrsim$1~GeV/nucleon are consistent given that we did not make stronger fine tuning than requiring the spectra be within the $\sim$few-percent data uncertainties, as described in Sec.~\ref{sec:tune}.
All configurations generally agree with the V1 data within the error bars down to energies $\sim$50~MeV nucleon$^{-1}$.
The baseline and S25/0/100 configurations are most discrepant for energies below this.
We do not discuss the baseline, because it is a result of the fit made by \citet{Cummings_2025}.
For the S25/0/100 configurations, the ratio is below the data because the Carbon spectrum is much higher than the other 3D/100/200 and S25/200 combinations.
Generally, for these other models the improvement for the low energy ratio comes from the attenuation of the denominator Carbon spectrum. 
Formally, the 3D/200 configuration is the closest to the data, but it is a setup with an idealised homogeneous gas distribution enhancing the energy losses compared to the data-driven S25 distribution.

We note that the attenuation of the low-energy Carbon spectrum is still not sufficient to raise the B/C ratio to obtain complete consistency with the V1 data.
This supports the contention of \citet{Cummings_2025} for a primary component contribution to the Boron spectrum.

Figure~\ref{fig2} shows the elemental spectra for Z=11--19 (left panel) and Z=20--28 (right panel).
The spectra for $\lesssim$100~MeV nucleon$^{-1}$ energies show sensitivity to the nearby ISM features of the models depending on their relative contributions via primary and secondary production channels.
For example, Iron is mostly primary and shows a very strong effect for the low-energy spectrum depending on source density hole size and gas model.
(This is similar to the case for C and O in Fig.~\ref{fig1a} that are also both mostly primary.)
Meanwhile, the low-energy spectra of the mixed-origin species (primary+secondary) reflect the relative contributions from the primary abundance as well as respective fragmentation cross sections \citep[][Appendix B.4]{Cummings_2025}.
Some of these elemental spectra (e.g., Vanadium) have only a very small primary abundance as determined from the 2D model fit, and it is only for the S25/0 configuration that the spectrum appears significantly different from the baseline case.

The calculated elemental ratios for He, O, and Fe for different setups are compared with V1 data in Figure~\ref{fig3}.
The V1 ratios are derived for each elemental combination by taking the intensity for each energy bin for whichever species has the least energy range coverage and then using power-law interpolation at the respective mid-bin energies\footnote{Following \citet{Cummings_2025} for given energy bin boundaries $E_{\rm max}$ and $E_{\rm min}$ this is ($E_{\rm min} + E_{\rm max}$)/2.} to obtain the intensities of the other species.
The uncertainties are likewise obtained using the same interpolation scheme combining the 1-$\sigma$ statistical uncertainties for each species \citep[following the prescription from the note for Table~C1 of][]{Cummings_2025}. 

For the left panel, which is showing the combinations with various hole sizes and the two gas models, the 2D fit (black solid) and S25/200 (cyan long dash dotted) are best matching to the data.
The short and long black-dashed lines are the 3D equivalent with 100 and 200~pc radius holes, respectively, in the source density distribution.
The dotted, short dash dotted, and long dash dotted are calculations made with the S25 gas model and no hole, 100~pc radius hole, and 200~pc radius hole in the source distribution, respectively.

The shapes of the curves are defined by the interplay of several processes, such as the ionisation energy losses ($\propto Z^2$), fragmentation (roughly $\propto A^{0.6}$, but is quite different at low energies), and injection.
The presence of the 100/200~pc radius hole in the source distribution leads to the rise of the ratios toward low energies.
This is mostly connected with the larger ionisation energy losses of the denominator species in the ratios, while the replenishment from the sources is absent.
While the local gas number density in the vicinity of the Solar system in the S25 model is lower than in the 2D/3D equivalent model, the diffusion coefficient is also lower by $\sim$20\%, which is a reflection of the effective large-scale gas distribution.
Both effects reduce the ionisation energy losses and fragmentation of the denominator species, thus the blue curves are always lower than the black curves except at very low energies $<$2--3~MeV nucleon$^{-1}$ 
for setups with the holes in the source distribution (3D/100/200, S25/100/200).
Note that the total fragmentation cross section for He is much smaller than that for the heavier species, O and Fe, see \citet[][Appendices~F and~G, Figs.~10 and~11]{2022ApJS..262...30P}.
For the latter, they are comparable even though the fragmentation cross section for Fe is larger than for O.

Very fast ionisation energy losses are dominating at low energies, which leads to a universal spectral shape for all primary species, and this explains the flattening in the black dashed curves seen at low energies.
The exact energy depends on $Z^2$; thus, for the He/O and He/Fe ratios it is observed at lower energies than for the O/Fe ratio.

The wavy shape of the ratios in the S25/0 setup appears due to a combination of the effects due to the ionisation energy losses, fragmentation, and injection.
The rise of the fraction for the low energies ($\lesssim$10~MeV nucleon$^{-1}$) is again due to the increased ionisation energy losses and fragmentation of the denominator species.
The right panel shows the effect of changing the source density hole size, where we have introduced an intermediate hole size of 150~pc (double-dot dashed) and extreme one of 400~pc (triple-dot dashed) for the S25 configuration to better illustrate the progression.
It is readily seen that the flattening shifts to even lower energies owing to the lower local gas number density.


\section{Discussion}
Our results show that low-energy CRs in the VLISM are jointly shaped by two distinct but physically linked forms of local inhomogeneity: the structured gas distribution, including the LB, and the corresponding deficit of recent CR accelerators within it.
Prior work has typically examined only one of these factors at a time.
Studies focused on source inhomogeneity using discrete, stochastic, or locally absent sources within a homogeneous ISM \citep[e.g.,][]{2021PhRvL.127n1101P}, have shown that the distance and age of nearby SN remnants can modulate the spectra of low-energy primaries.
Conversely, other investigations isolating effects of a local underdensity in the gas while retaining a smooth, Galaxy-wide source population \citep[e.g.,][]{2002A&A...381..539D} have mainly looked at effects on secondary CRs without considering energy losses.
The picture that we have now of the LB and its surroundings suggests that it was excavated by SN explosions, leading to an interior rarified ISM and a highly structured distribution of neutral and molecular gas at its boundary \citep{Zucker_2022}. 
The SN remnants have long since dispersed with the consequence that the space within the LB is also deficient in terms of CR injection.
Our study provides the first explicit assessment of how these effects can act together to shape the low-energy CR spectra in the ISM.

Our modelling reveals that considering these inhomogeneities individually produces spectral behaviours that are inconsistent with the V1 measurements.
When only the realistic 3D gas structure is included but the source distribution remains spatially uniform, the low-energy spectra are systematically too high.
The depressed gas density near the Solar system reduces ionisation losses for primaries, allowing an excess of low-energy particles to persist.
Alternatively, imposing a local deficit of sources within a homogeneous gas distribution produces spectra that fall too steeply unless the hole radius is tuned to compensate for the incorrect loss environment.
In both cases, isolating gas or source structure forces the model to compensate by distorting the low-energy behaviour in ways that do not reflect the actual physical conditions of the nearby Galactic neighbourhood.

When the two forms of structure are combined, a consistent and physically motivated picture emerges.
The S25 gas distribution captures the LB’s underdense interior and its irregular boundary of enhanced gas density, while the introduction of a $\sim$150--200~pc source region deficit about the Solar system models the absence of a recent and sufficiently strong hadronic CR injector inside the cavity.
Together, these components naturally reproduce the observed low-energy softening: particles originate at or just beyond the LB boundary, traverse regions of both reduced and elevated density, and accumulate energy losses in a manner consistent with the measured Voyager spectra.
In this picture, the attenuation of primaries is neither dominated solely by propagation distance (as in a homogeneous ISM) nor artificially suppressed by a uniformly low density.
Rather, it arises from the interaction between both structured CR injection and losses.

The statistical likelihood of such a source deficit deserves consideration.
Adopting canonical values for an order-of-magnitude expectation, we take a Galactic CR source rate of $\sim$1 per century and an effective ``active'' time window of $\Delta t \sim 10^5$ yr over which a single source can significantly influence the local sub-GeV/n nuclei spectra \citep[e.g.,][]{1999ApJ...523L..61W}.
This implies $\sim$$10^3$ potentially relevant sources Galaxy-wide over $\Delta t$.
If we approximate the spatial distribution as uniform over a star-forming disc of radius $R_{\rm disc}\sim15$~kpc and half-thickness $Z_{\rm disc}\sim0.1$~kpc (volume $V_{\rm disc} = \pi R_{\rm disc}^2 (2Z_{\rm disc})\approx 1.4\times 10^2~{\rm kpc}^3$), then the expected number inside a sphere of radius $R$ is $\lambda(R)\approx (10^3)\,V_{\rm sph}(R)/V_{\rm disc}$, with $V_{\rm sph}=4\pi R^3/3$.
For $R=150$ pc, $V_{\rm sph}\approx 1.4\times 10^{-2}~{\rm kpc}^3$, giving $\lambda\simeq 0.10$ and hence $P(0)=e^{-\lambda}\simeq 0.90$.
For $R=200$~pc, $V_{\rm sph}\approx 3.4\times 10^{-2}~{\rm kpc}^3$, giving $\lambda\simeq 0.24$ and $P(0)\simeq 0.79$.
Even for a higher rate of 3 per century, probabilities remain significant at 0.74 and 0.49, respectively.  
Thus, from Poisson statistics alone, a local deficit of effective CR sources is plausible.

This statistical picture is supported by the observations of the distribution of Wolf--Rayet and OB stars.
These stars are candidates for CR acceleration throughout their lives via powerful winds, in addition to being the progenitors of SNe \citep[e.g.,][]{2023arXiv231015442M}.
Crucially, recent 3D maps show that the star-forming regions containing these active and future CR sources are situated on the surface of the LB, expanding outward from its interior \citep{Zucker_2022}.
The fact that these massive stars are located at the boundary of the LB provides physical backing for our conclusion of a source-deficient volume within it.

While the evidence points to a lack of recent and future source activity within the LB, this must be reconciled with the known population of nearby pulsars.
The ATNF Pulsar Catalogue lists a couple dozen objects within $\sim$200~pc.
However, our finding applies specifically to hadronic CR sources.
While pulsars are confirmed electron-positron accelerators, their capacity to accelerate nuclei to significant energies is less certain, although mechanisms for their production and release from the confining pulsar wind nebula have been investigated \citep[e.g.,][]{2002APh....16..397B,2004A&A...423..405B,2004NuPhS.136..185B}.

Furthermore, due to large proper motions imparted by natal kicks, the current location of a pulsar can be far removed from the site of its parent SN.
Given the age of the LB, the SN remnants that created it and any associated pulsars have long since dispersed or travelled significant distances.
The presence of older, displaced pulsars is therefore not inconsistent with a local deficit of recent, powerful CR acceleration events.
Consequently, our result does not contradict the existence of these pulsars but implies that either (i) these specific nearby objects are not significant contributors to the hadronic CR flux, or (ii) their contribution is subdominant to that from more distant sources, at least for the low-energy particles measured by V1.

Meanwhile, the behaviour of secondary-dominated species is a further clue.
Because the Li–Be–B production samples larger effective volumes, their fluxes remain relatively insensitive to localised features in either the gas or the source distribution, and their ratios vary only modestly across configurations that we have considered.
However, as noted before when discussing Fig.~\ref{fig2}, \citet{Cummings_2025} find evidence for a low-energy primary Boron contribution that would complicate interpretation of the B/C ratio in this picture.
For this scenario (called 4022c in their work), we used the S25 gas distribution and followed the tuning procedure (Sec.~\ref{sec:tune}) to normalise the configuration, and predict the corresponding B/C for hole sizes 0 and 150~pc.
The results are shown in the left panel of Fig.~\ref{fig4} together with those for the corresponding 4022a (pure secondary B) configuration with the same hole sizes, as used earlier in this work.
The energy range for this figure is reduced compared to earlier, because the models with primary Boron were fit only using the V1 data.
The difference between configurations that include a primary Boron contribution and those that do not is as large, if not larger than, those associated with the inhomogeneous source and ISM distributions investigated already. 
Using this secondary/primary ratio both as a diagnostic for the LISM conditions and global propagation parameters may not be as reliable as conventionally taken.

Figure~\ref{fig4} (right panel) shows the Be/O ratio that we suggest may be a more useful constraint for propagation scenarios.
Data and uncertainties were obtained using the same interpolation procedure as employed for Fig.~\ref{fig3}.
Beryllium is a pure secondary, and the Be/O ratio is fairly insensitive to the low-energy effects seen for B/C with all scenarios consistent with the V1 data.
At higher energies, the configurations with/without primary Boron may be distinguishable using higher-energy data from ACE-CRIS\footnote{https://izw1.caltech.edu/ACE/ASC/level2/lvl2DATA\_CRIS.html} and AMS-02 \citep{2021PhR...894....1A}.
This issue requires accurate Solar modulation models, e.g., HelMod \citep{2020ApJS..250...27B}, and will be addressed in a follow-up paper.

\section{Summary}
We have investigated the impact of realistic local gas structure and source distribution inhomogeneity on low-energy CR spectra using the \GP\ modelling framework.
Our models include inhomogeneous source density and high-resolution ISM models.
Comparison of them with V1 measurements in the VLISM shows that the combination of an inhomogeneous ISM and CR source distribution is necessary to reproduce the data.
Neither alone is sufficient.
Indeed, only considering one factor will lead to incorrect conclusions about the physical conditions shaping the low-energy CR spectra.

The best agreement with V1 data is when there is no dominant/recent source of CR nuclei within $\sim$150--200~pc (for the S25 gas reconstruction).
For this scenario, the low-energy CR fluxes are coming from sources just beyond the LB boundary.
While our modelling assumes a simplified source distribution outside this boundary, the fundamental conclusion that a combination of local gas structure and a local source deficit is necessary remains robust.
Future work incorporating specific, powerful nearby catalogue sources like Vela and Geminga could further refine the exact parameters of this local void.
Interplay between reduced energy losses in the underdense interior, enhanced losses in the region outside, and the proximity of the sources produce the observed spectra.
The secondary-dominated species remain largely unaffected by these localised features; consequently, the primary spectra carry the strongest imprint of nearby ISM structure.

The conclusions by \citet{Cummings_2025} about the significant primary contribution to the observed Boron spectrum at low energies are supported by our work.
This raises doubts in the widely accepted paradigm about Boron being 100\% secondary in CRs and the derivation of propagation parameters from the B/C or B/O ratio. 

Our study highlights the importance of incorporating the realistic structure of the nearby Galactic neighbourhood when interpreting low-energy CR data, or using such data to infer global Galactic propagation parameters.
Both the gas distribution and the recent source history of the nearby ISM must be accounted for to obtain unbiased constraints and to correctly interpret the V1 data.

\begin{acknowledgements}
  This work is partially funded via NASA grants 80NM0018F0577, 80NSSC22K0477, 80NSSC22K0718, and 80NSSC23K0169.
\end{acknowledgements}

\bibliography{lowecr}
  
\end{document}